\documentclass[11pt]{article}
\begin{document}

\title{ 
{\bf Large Gauge Invariance and Finite Temperature}}
\author{Ashok Das \\
\\
Department of Physics and Astronomy, \\
University of Rochester,\\
Rochester, New York, 14627.}
\date{}
\maketitle

\begin{abstract}

In this talk, I  summarize the status of our understanding of the
puzzle of large gauge invariance at finite temperature.

\end{abstract}

\vfill\eject
\section{Introduction:}

Gauge theories are beautiful theories which describe physical forces
in a natural manner and because of their rich structure, the study of
gauge theories at finite temperature \cite{l1} is quite interesting in itself.
However, to avoid getting into technicalities, we will
not discuss the intricacies of such theories either at zero
temperature or at finite temperature. Rather, we would study a
particular puzzle that arises when a fermion interacts with an
external gauge field in odd space-time dimensions.

To motivate, let us note that gauge invariance is realized as an
internal symmetry in quantum mechanical systems. Consequently, we do
not expect a macroscopic external surrounding, such as a heat bath, to
modify gauge invariance. This is more or less what is also found by
explicit computations at finite temperature, namely, that gauge
invariance and Ward identities continue to hold even at finite
temperature  \cite{l2}. This is certainly the case when one is talking about
small gauge transformations for which the parameters of transformation
vanish at infinity.

However, there is a second class of gauge transformations, commonly known
as large gauge transformations, where the parameters do not vanish at
infinity. In odd space-time dimensions, invariance under large gauge
transformations leads to some interesting features in physical
theories. Let us note that, in odd space-time dimensions, one can, in
addition to the usual Maxwell (Yang-Mils) term, have a topological
term in the gauge Lagrangian known as the Chern-Simons term.  For
example,  in $2+1$ dimensions, a fermion interacting with a
non-Abelian gauge field can be described by a Lagrangian density of
the form
\begin{eqnarray}
{\cal L} & = & {\cal L}_{\rm gauge} -  m{\cal L}_{\rm CS} + {\cal
 L}_{\rm fermion}\nonumber\\
 & = & {1\over 2}\,{\rm tr}\,F_{\mu\nu}F^{\mu\nu} -
 m\epsilon^{\mu\nu\lambda}\,{\rm
 tr}\,A_{\mu}(\partial_{\nu}A_{\lambda}+{2g\over 3}A_{\nu}A_{\lambda}) +
 \overline{\psi}(\gamma^{\mu}(i\partial_{\mu}-gA_{\mu})-M)\psi\label{g1}
\end{eqnarray}
where $m$ is a mass parameter, $A_{\mu}$  a matrix valued non-Abelian
gauge  field in a given representation and \lq\lq tr'' stands for the
matrix  trace. The first term, on the right hand side, is the usual
Yang-Mills term while the second is known as the Chern-Simons term which
exists only in odd space-time dimensions. It is a topological term
(since it does not involve the metric) and, in the presence of a
Yang-Mils term, its effect is to provide a gauge invariant mass term
to the gauge fields. Consequently, such a term is also known as a
topological mass term  \cite{l3}. (Such a term also breaks various
discrete symmetries, but we will not get into that.)

Under a gauge transformation of the form
\begin{eqnarray}
\psi & \rightarrow & U^{-1}\,\psi\nonumber\\
A_{\mu} & \rightarrow & U^{-1}\,A_{\mu}\,U - {i\over
g}\,U^{-1}\,\partial_{\mu}U\label{g2}
\end{eqnarray}
it is straightforward to check that both the Yang-Mills and the
fermion terms are invariant, while the Chern-Simons term changes by a
total divergence leading to
\begin{equation}
S = \int d^{3}x\,{\cal L} \rightarrow S + {4\pi m\over g^{2}}\,2i\pi
W\label{g3}
\end{equation}
where 
\begin{equation}
W = {1\over 24\pi^{2}}\int d^{3}x\,\epsilon^{\mu\nu\lambda}\,{\rm
tr}\partial_{\mu}UU^{-1}\partial_{\nu}UU^{-1}\partial_{\lambda}UU^{-1}
\label{g4}
\end{equation}
is known as the winding number for the gauge transformation. It is a
topological quantity which is an integer and which groups all gauge
transformations into topologically distinct classes. Basically, it
counts how many times the gauge transformations wrap around the
sphere. For small gauge transformations, the winding number vanishes
since the gauge transformations vanish at infinity whereas
non-vanishing winding numbers give rise to large gauge
transformations.

Let us note from eq. (\ref{g3}) that even though the action is not
invariant under a large gauge transformation, if $m$ is quantized in
units of ${g^{2}\over 4\pi}$, the change in the action would be a
multiple of $2i\pi$ and, consequently, the path integral would be
invariant under a large gauge transformation. Thus, we have the
constraint coming from the consistency of the theory that the
coefficient of the Chern-Simons term must be quantized in units of
${g^{2}\over 4\pi}$. (From an
operator point of view, such a condition arises from an analysis of
the dimensionality of the Hilbert space.) 

We have derived the quantization of the CS coefficient from an
analysis of the large gauge invariance of the tree level 
action and we have to worry if the quantum corrections can change
the behavior of the theory. At zero temperature, an analysis of the
quantum corrections shows that the theory continues to be well defined
with the tree level quantization of the Chern-Simons coefficient
provided the number of fermion flavors is even. The even number of
fermion flavors is also necessary for a global anomaly of the theory to
vanish and so, everything is well understood at zero
temperature. (Namely, the quantum corrections, with an even number of
fermion flavors shift the tree level integer value to another, thereby
maintaining large gauge invariance.)

At finite temperature, however, the situation appears to change
drastically. Namely, the fermions induce a temperature dependent
Chern-Simons term leading to \cite{l4}
\begin{equation}
m \rightarrow m -{g^{2}\over 4\pi}\,{MN_{f}\over 2|M|}\,\tanh
{\beta|M|\over 2}\label{g5}
\end{equation}
Here, $N_{f}$ is the number of fermion flavors and $\beta = {1\over
kT}$.  This shows that, at
zero temperature ($\beta\rightarrow\infty$), $m$ changes by an integer
(in units of $g^{2}/4\pi$) for an even number of flavors. However, at
finite temperature, the CS coefficient becomes a continuous function
of temperature 
and, consequently, it is clear that it can no longer be an integer for
arbitrary values of the temperature, as is required for large gauge
invariance.  It seems, therefore, that temperature
would lead to a breaking of large gauge invariance in such a
system. This, on the other hand, is completely counter intuitive
considering that temperature should have no direct influence on gauge
invariance of the theory. As a result, we are left with a puzzle,
whose resolution, as we will see, is quite interesting.

\section{C-S Theory in $0+1$ Dimension:}

The $2+1$ dimensional theory described in the previous section is
quite complicated to carry out higher order calculations. On the other
hand, as we have noted, Chern-Simons terms can exist in odd space-time
dimensions. Consequently, let us try to understand this puzzle of
large gauge invariance in a simple quantum mechanical theory. Let us
consider a simple theory of an interacting massive fermion with an
Abelian gauge field in $0+1$ dimension described by  \cite{l5,l6}
\begin{equation}
L = \overline{\psi}_{j}(i\partial_{t} - A - M)\psi_{j} - \kappa
A\label{g6}
\end{equation}
Here, $j=1,2,\cdots,N_{f}$ labels the fermion flavors. There are
several things to note from this. First, we are considering an Abelian
gauge field for simplicity. Second, in this simple model, the gauge
field has no dynamics (in $0+1$ dimension the field strength is zero)
and, therefore, we do not have to get into the intricacies of gauge
theories. There is no Dirac matrix in $0+1$ dimension as well making
the fermion part of the theory quite simple as well. And, finally, the
Chern-Simons term, in this case, is a linear field so that we can, in
fact, think of the gauge field as an auxiliary field. (Note that we
have set the coupling constant to unity for simplicity.)

In spite of the simplicity of this theory, it displays a rich
structure including all the properties of the $2+1$ dimensional
theory that we have discussed earlier. For example, let us note that
under a gauge transformation
\begin{equation}
\psi_{j}\rightarrow e^{-i\lambda(t)}\psi_{j},\qquad A\rightarrow A +
\partial_{t}\lambda(t)\label{g7}
\end{equation}
the fermion part of the Lagrangian is invariant, but the Chern-Simons
term changes by a total derivative giving
\begin{equation}
S = \int dt\,L \rightarrow S -2\pi\kappa N\label{g8}
\end{equation}
where 
\begin{equation}
N = {1\over 2\pi} \int dt\,\partial_{t}\lambda(t)\label{g9}
\end{equation}
is the winding number and is an integer which vanishes for small gauge
transformations. Let us note that a large gauge transformation can have
a parametric form of the form, say,
\begin{equation}
\lambda(t) = -iN\,\log\left({1+it\over 1-it}\right)\label{g10}
\end{equation}
The fact that $N$ has to be an integer can be easily seen to arise
from the requirement of single-valuedness for the fermion field. Once
again, in light of our earlier discussion, it is clear from
eq. (\ref{g8}) that the theory is meaningful only if $\kappa$, the
coefficient of the Chern-Simons term, is an integer.

Let us assume, for simplicity, that $M>0$ and compute the correction
to the photon one-point function arising from the fermion loop at zero
temperature.
\begin{equation}
iI_{1} = -(-i)N_{f}\,\int {dk\over 2\pi}\,{i(k+M)\over
k^{2}-M^{2}+i\epsilon} = {iN_{f}\over 2}\label{g11}
\end{equation}
This shows that, as a result of the quantum correction, the
coefficient of the Chern-Simons term would change as
\[
\kappa\rightarrow \kappa - {N_{f}\over 2}
\]
As in $2+1$ dimensions, it is clear that the coefficient of the
Chern-Simons term would continue to be quantized and large gauge
invariance would hold if the number of fermion flavors is even. At
zero temperature, we can also calculate the higher point functions due
to the fermions in the theory and they all vanish. This has a simple
explanation following from the small gauge invariance of the
theory  \cite{l7}. Namely, suppose we had a nonzero two point function, then, it
would imply a quadratic term in the effective action of the form
\begin{equation}
\Gamma_{2} = {1\over 2} \int
dt_{1}\,dt_{2}\,A(t_{1})F(t_{1}-t_{2})A(t_{2})\label{g12}
\end{equation}
Furthermore, invariance under a small gauge transformation would imply
\begin{equation}
\delta\Gamma_{2} = -\int
dt_{1}\,dt_{2}\,\lambda(t_{1})\partial_{t_{1}}F(t_{1}-t_{2})A(t_{2}) =
0\label{g13}
\end{equation}
The solution to this equation is that $F=0$ so that there cannot be a
quadratic term in the effective action which would be local and yet be
invariant under small gauge transformations. A similar analysis would
show that small gauge invariance does not allow any higher point
function to exist at zero temperature. 

Let us also note that eq. (\ref{g13}) has another solution, namely,
\[
F(t_{1}-t_{2}) = {\rm constant}
\]
In such a case, however, the quadratic action becomes non-extensive,
namely, it is the square of an action. We do not expect such terms to
arise at zero temperature and hence the constant has to vanish for
vanishing temperature. As we will see next, the constant does not have
to vanish at finite temperature and we can have non-vanishing higher
point functions implying a non-extensive structure of the effective
action. 

The fermion propagator at finite temperature (in the real time
formalism) has the form  \cite{l1}
\begin{eqnarray}
S(p) & = & (p+M)\left({i\over p^{2}-M^{2}+i\epsilon} -
2\pi n_{F}(|p|)\delta(p^{2}-M^{2})\right)\nonumber\\
 & = & {i\over p-M+i\epsilon} - 2\pi n_{F}(M)\delta(p-M)\label{g14}
\end{eqnarray}
and the structure of the effective action can be studied in the
momentum space in a straightforward manner. However, in this simple
model,  it is much easier to analyze the amplitudes in the coordinate
space.  Let us note that the coordinate space structure of the fermion
propagator is quite simple, namely,
\begin{equation}
S(t) = \int {dp\over 2\pi}\,e^{-ipt}\,\left({i\over p-M+i\epsilon} -
2\pi n_{F}(M)\delta(p-M)\right) =
(\theta(t)-n_{F}(M))e^{-iMt}\label{g15}
\end{equation}
In fact, the calculation of the one point function is trivial now
\begin{equation}
iI_{1} = -(-i)N_{f}S(0) = {iN_{f}\over 2}\,\tanh {\beta M\over
2}\label{g16}
\end{equation}
This shows that the behavior of this theory is completely parallel to
the $2+1$ dimensional theory in that, it would suggest
\[
\kappa\rightarrow \kappa -{N_{f}\over 2}\,\tanh {\beta M\over 2}
\]
and it would appear that large gauge invariance would not hold at
finite temperature.

Let us next calculate the two point function at finite temperature.
\begin{eqnarray}
iI_{2} & = & -(-i)^{2}\,{N_{f}\over
2!}\,S(t_{1}-t_{2})S(t_{2}-t_{1})\nonumber\\
 & = & -{N_{f}\over 2}\,n_{F}(M)(1-n_{F}(M))\nonumber\\
 & = & -{N_{f}\over 8}\, {\rm sech}^{2} {\beta M\over 2} = {1\over
2}\,{1\over 2!}\,{i\over\beta}\,{\partial(iI_{1})\over \partial M}\label{g17}
\end{eqnarray}
This shows that the two point function is a constant as we had noted
earlier  implying that the quadratic term in the effective action
would be non-extensive.

Similarly, we can also calculate the three point function trivially
and it has the form
\begin{equation}
iI_{3} = {iN_{f}\over 24}\,\tanh {\beta M\over 2}\,{\rm sech}^{2} {\beta
M\over 2} = {1\over 2}\,{1\over
3!}\,\left({i\over\beta}\right)^{2}\,{\partial^{2}(iI_{1})\over
\partial M^{2}}\label{g18}
\end{equation}
In fact, all the higher point functions can be worked out in a
systematic manner. But, let us observe a simple method of computation
for these. We note that because of the gauge invariance (Ward
identity), the amplitudes cannot depend on the external time
coordinates as is clear from the  calculations of the lower point
functions. Therefore, we can always simplify the calculation by
choosing a particular time ordering convenient to us. Second, since we
are evaluating a loop diagram (a fermion loop) the initial and the
final time coordinates are the same and, consequently, the phase
factors in the propagator (\ref{g15}) drop out. Therefore, let us
define a simplified propagator without the phase factor as
\begin{equation}
\widetilde{S}(t) = \theta(t) - n_{F}(M)\label{g19}
\end{equation}
so that we have
\begin{equation}
\widetilde{S}(t>0) = 1 - n_{F}(M),\qquad S(t<0) = -
n_{F}(M)\label{g20}
\end{equation}
Then, it is clear that with the choice of the time ordering,
$t_{1}>t_{2}$, we can write
\begin{eqnarray}
{\partial\widetilde{S}(t_{1}-t_{2})\over \partial M} & = & -\beta
\widetilde{S}(t_{1}-t_{3})\widetilde{S}(t_{3}-t_{2})\qquad
t_{1}>t_{2}>t_{3}\nonumber\\
{\partial\widetilde{S}(t_{2}-t_{1})\over \partial M} & = & -\beta
\widetilde{S}(t_{2}-t_{3})\widetilde{S}(t_{3}-t_{1})\qquad
t_{1}>t_{2}>t_{3}\label{g21}
\end{eqnarray}

In other words, this shows that differentiation of a fermionic
propagator with respect to the mass of the fermion is equivalent to
introducing an external photon vertex (and, therefore, another fermion
propagator as well) up to constants. This is the analogue of the Ward identity
in QED in four dimensions except that it is much simpler. From this
relation, it is clear that if we take a $n$-point function and
differentiate this with respect to the fermion mass, then, that is
equivalent to adding another external photon vertex in all possible
positions. Namely, it should give us the $(n+1)$-point function up to
constants. Working out the details, we have,
\begin{equation}
{\partial I_{n}\over \partial M} = -i\beta(n+1)I_{n+1}\label{g22}
\end{equation}
Therefore, the $(n+1)$-point function is related to the $n$-point
function recursively and, consequently, all the amplitudes are
related to the one point function which we have already
calculated. (Incidentally, this is already reflected in
eqs. (\ref{g17},\ref{g18})). 

With this, we can now determine the full effective action of the
theory at finite temperature to be
\begin{eqnarray}
\Gamma & = & -i\,\sum_{n}\,a^{n}\,(iI_{n})\nonumber\\
 & = & -{i\beta N_{f}\over 2}\,\sum_{n}\,{(ia/\beta)^{n}\over
 n!}\,\left({\partial\over \partial M}\right)^{n-1}\,\tanh {\beta
 M\over 2}\nonumber\\
 & = & -iN_{f}\,\log \left(\cos {a\over 2} + i \tanh {\beta M\over
 2}\,\sin {a\over 2}\right)\label{g23}
\end{eqnarray}
where we have defined
\begin{equation}
a = \int dt\,A(t)\label{g24}
\end{equation}

There are several things to note from this result. First of all, the
higher point functions are no longer vanishing at finite temperature
and give rise to a non-extensive structure of the effective
action. More importantly, when we include all the higher point
functions, the complete effective action is invariant under large
gauge transformations, namely, under
\begin{equation}
a\rightarrow a + 2\pi N\label{g25}
\end{equation}
the effective action changes as 
\begin{equation}
\Gamma\rightarrow \Gamma + N N_{f}\pi\label{g26}
\end{equation}
which leaves the path integral invariant for an even number of fermion
flavors. This clarifies the puzzle of large gauge invariance at finite
temperature in this model. Namely, when we are talking about
large changes (large gauge transformations), we cannot ignore higher
order terms if they exist. This may provide a resolution to the large
gauge invariance puzzle in the $2+1$ dimensional theory as well. 

\section{Exact Result:}

In the earlier section, we discussed a perturbative method of
calculating the effective action at finite temperature which clarified
the puzzle of large gauge invariance. However, this quantum mechanical
model is simple enough that we can also evaluate the effective action
directly and, therefore, it is worth asking how  the perturbative
calculations compare with the exact result.

The exact evaluation of the effective action can be done easily using
the imaginary time formalism. But, first, let us note that the
fermionic part of the Lagrangian in eq. (\ref{g6}) has the form
\begin{equation}
L_{f} = \overline{\psi}(i\partial_{t} - A - M)\psi\label{g27}
\end{equation}
where we have suppressed the fermion flavor index for simplicity. Let
us note that if we make a field redefinition of the form
\begin{equation}
\psi(t) = e^{-i\int_{0}^{t} dt'\,A(t')}\,\tilde{\psi}(t)\label{g28}
\end{equation}
then, the fermionic part of the Lagrangian becomes free, namely,
\begin{equation}
L_{f} = \overline{\tilde{\psi}}\,(i\partial_{t} -
M)\tilde{\psi}\label{g29}
\end{equation}

This is a free theory and, therefore, the path integral can be easily
evaluated. However, we have to remember that the field redefinition in
(\ref{g28})  changes the periodicity condition for the fermion
fields. Since the original fermion field was expected to satisfy 
anti-periodicity 
\[
\psi(\beta) = - \psi(0)
\]
it follows now that the new fields must satisfy
\begin{equation}
\tilde{\psi}(\beta) = - e^{-ia}\,\tilde{\psi}(0)\label{g30}
\end{equation}
Consequently, the path integral for the free theory (\ref{g29}) has to
be evaluated subject to the periodicity condition of (\ref{g30}).

Although the periodicity condition (\ref{g30}) appears to be
complicated, it is well known that this can be absorbed by
introducing a chemical potential  \cite{l1}, in the present case, of the form
\begin{equation}
\mu = {ia\over \beta}\label{g31}
\end{equation}
With the addition of this chemical potential, the path integral can be
evaluated  subject to the usual anti-periodicity condition. The
effective action can now be easily determined
\begin{eqnarray}
\Gamma & = & -i\,\log \left({\det
(i\partial_{t}-M+{ia\over\beta})\over
(i\partial_{t}-M)}\right)^{N_{f}}\nonumber\\
 & = & -iN_{f}\,\log \left({\cosh {\beta\over 2}(M-{ia\over \beta})\over
\cosh {\beta M\over 2}}\right)\nonumber\\
 & = & -iN_{f}\,\log \left(\cos {a\over 2} + i\tanh {\beta M\over
2}\,\sin {a\over 2}\right)\label{g32}
\end{eqnarray}
which coincides with the perturbative result of eq. (\ref{g24}).

\section{Large Gauge Ward Identity:}

It is clear from the above analysis that, to see if large gauge
invariance is restored, we have to look at the complete effective
action. In the $0+1$ dimensional model, it was tedious, but we can
derive the effective action in closed form which allows us to analyze
the question of large gauge invariance. On the other hand, in the
theory of interest, namely, the $2+1$ dimensional Chern-Simons theory,
we do not expect to be able to evaluate the effective action in a
closed form. Consequently, we must look for an alternate way to
analyze the question of large gauge invariance in a more realistic
model. One such possible method may be to derive a Ward identity for
large gauge invariance which will relate different amplitudes much
like the Ward identity for small gauge invariance does. In such a
case, even if we cannot obtain the effective action in a closed form,
we can at least check if the large gauge Ward identity holds
perturbatively.

It turns out that the large gauge Ward identities are highly
nonlinear  \cite{l8}, as we would expect. Hence, looking for them within the
context of the effective action is extremely hard (although it can be
done). Rather, it is much simpler to look at the large gauge Ward
identities in terms of the exponential of the effective action. Let us
define 
\begin{equation}
\Gamma(a) = - i \log W(a)\label{9}
\end{equation}
Namely, we are interested in looking at the
exponential of the effective action ({\it i.e.} up to a factor of $i$,
$W$ is the basic determinant that would arise from integrating out the
fermion
field). We will restrict ourselves to a single flavor
of massive fermions. The
advantage of studying $W(a)$ as opposed to the effective action lies
in the fact that, in order for $\Gamma(a)$ to have the right
transformation
properties  under a large gauge transformation, $W(a)$ simply has to be
quasi-periodic. Consequently, from the study of harmonic oscillator (as
well as Floquet theory), we see that $W(a)$ has to satisfy a simple
equation of the form
\begin{equation}
{\partial^{2}W(a)\over \partial a^{2}} + \nu^{2}\,W(a) = g\label{10}
\end{equation}
where $\nu$ and $g$ are parameters to be determined from the
theory. In particular, let us note that the constant $g$ can depend on
parameters of the theory such as temperature whereas we expect the
parameter
$\nu$, also known as the characteristic exponent, to be independent of
temperature and equal to an odd half integer for a fermionic
mode. However,  all these properties should
automatically result from the structure of the theory. Let us also
note here that the relation (\ref{10}) is simply the equation for a
forced oscillator whose solution has the general form
\begin{equation}
W(a) = \frac{g}{\nu^2} + A\cos(\nu a + \delta) = \frac{g}{\nu^2} +
\alpha_{1}\cos\nu a +
\alpha_{2}\sin\nu a\label{10''}
\end{equation}
The constants $\alpha_{1}$ and $\alpha_{2}$ appearing in the solution
can again be determined from the theory. Namely, from the relation
between $W(a)$ and $\Gamma(a)$, we recognize that we can identify
\begin{eqnarray}
\nu^{2}\alpha_{1} & = & - \left.{\partial^{2}W\over \partial
a^{2}}\right|_{a=0} = \left.\left(\left({\partial\Gamma\over \partial
a}\right)^{2} - i\,{\partial^{2}\Gamma\over \partial
a^{2}}\right)\right|_{a=0}\nonumber\\
\noalign{\kern 4pt}%
\nu\alpha_{2} & = & \left.{\partial W\over \partial a}\right|_{a=0} =
i\,\left.{\partial\Gamma\over \partial a}\right|_{a=0}\label{10'}
\end{eqnarray}
From the general properties of the fermion theories we have
discussed, we intuitively expect $g=0$. However, these should really
follow  from the
structure of the theory and they do, as we will show shortly.

The identity (\ref{10}) is a linear relation as opposed to the Ward
identity  in terms of the effective action. In fact, rewriting this in terms
of the effective action (using eq. (\ref{9})), we have
\begin{equation}
{\partial^{2}\Gamma(a)\over \partial a^{2}} =
 i\left(\nu^{2}-\left({\partial\Gamma(a)\over \partial
a}\right)^{2}\right) - i g\,e^{- i\Gamma(a)}\label{11}
\end{equation}
So, let us investigate this a little
bit  more in detail. We know
that the fermion mass term breaks parity and, consequently, the
radiative corrections would generate a Chern-Simons term, namely, in
this theory, we expect the one-point function to be
nonzero. Consequently, by taking  derivative of eq. (\ref{11}) (as
well as remembering that $\Gamma(a=0)=0$), we determine (The
superscript represents the number of flavors.)
\begin{eqnarray}
(\nu^{(1)})^{2} & = & \left[\left({\partial\Gamma^{(1)}\over
\partial
a}\right)^{2} - 3i\,{\partial^{2}\Gamma^{(1)}\over \partial a^{2}}
- \left({\partial\Gamma^{(1)}\over \partial
a}\right)^{-1}\left({\partial^{3}\Gamma^{(1)}\over \partial
a^{3}}\right)\right]_{a=0}\nonumber\\
\noalign{\kern 4pt}%
g^{(1)} & = & -\left[2i\,{\partial^{2}\Gamma^{(1)}\over \partial
a^{2}} + \left({\partial\Gamma^{(1)}\over \partial
a}\right)^{-1}\left({\partial^{3}\Gamma^{(1)}\over \partial
a^{3}}\right)\right]_{a=0}\label{12}
\end{eqnarray}
This is quite interesting, for it says that the two parameters in
eq. (\ref{10}) or (\ref{11}) can be determined from a perturbative
calculation. Let us note here some of the perturbative results in this
theory, namely,
\begin{eqnarray}
\left.{\partial\Gamma^{(1)}\over \partial a}\right|_{a=0} & = &
{1\over 2}\,\tanh {\beta M\over 2}\nonumber\\
\noalign{\kern 4pt}%
\left.{\partial^{2}\Gamma^{(1)}\over \partial a^{2}}\right|_{a=0}
& = & {i\over 4}\,{\rm sech}^{2} {\beta M\over 2}\nonumber\\
\noalign{\kern 4pt}%
\left.{\partial^{3}\Gamma^{(1)}\over \partial a^{3}}\right|_{a=0}
& = & {1\over 4}\,\tanh {\beta M\over 2}\,{\rm sech}^{2} {\beta M\over
2}
\end{eqnarray}
Using these, we immediately determine from eq. (\ref{12}) that
\begin{equation}
(\nu^{(1)})^{2} = {1\over 4},\quad\quad g^{(1)} = 0\label{13}
\end{equation}
so that the equation (\ref{11}) leads to the large gauge Ward identity
for a single fermion theory of the form,
\begin{equation}
{\partial^{2}\Gamma^{(1)}\over \partial a^{2}} = i\left({1\over 4} -
 \left({\partial\Gamma^{(1)}\over \partial a}\right)^{2}\right)
\end{equation} 
Furthermore, we determine now from eq. (\ref{10'})
\begin{equation}
\alpha_{1}^{(1)} = 1,\quad\quad \alpha_{2}^{(1)} = \pm i\,\tanh
{\beta M \over 2}
\end{equation}
The two signs in of $\alpha_{2}^{(1)}$ simply corresponds to the two
possible signs of $\nu^{(1)}$. With this then, we can solve for
$W(a)$ in  the single flavor fermion  theory and we have (independent
of the sign of $\nu^{(1)}$)
\begin{equation}
W_{f}^{(1)}(a) = \cos{a\over 2} + i\,\tanh{\beta M\over 2}\sin{a\over
2}\label{13'}
\end{equation}
which can be compared with eq. (\ref{g32}). For $N_{f}$ flavors,
similarly, we can determine the Ward identity to be
\begin{equation}
{\partial\Gamma^{(N_{f})}\over \partial a^{2}} = iN_{f}\left({1\over
4}-{1\over N_{f}^{2}}\left({\partial\Gamma^{(N_{f})}\over \partial
a^{2}}\right)^{2}\right)
\end{equation}
where the nonlinearity of the Ward identity is manifest.

Similarly, we can determine the large gauge Ward identity for scalar
theories as well as supersymmetric theories, but we will not go into
the details of this.

\section{Back to $2+1$ dimensions:}

The analysis of various $0+1$ dimensional models shows \cite{bar} how the large
gauge invariance puzzle gets resolved. A crucial feature in this was
the existence of non-extensive higher order terms in the effective
action. To further understand this feature, we have also studied
the effective action for a fermion interacting with an external gauge
field in $1+1$ dimensions \cite{daS}. In that case, the effective action is
non-local, as it should be at finite temperature, but 
extensive. Furthermore, the effective action shows
non-analyticity, as one would expect at finite temperature, unlike the
$0+1$ dimensional model, where one does not expect any
non-analyticity.

Following the results of the
$0+1$ dimensional model, it was shown  \cite{l9} that, for the special
choice of the gauge field backgrounds where $A_{0}=A_{0}(t)$ and
$\vec{A}= \vec{A}(\vec{x})$, the parity violating part of the
effective action of a  fermion interacting with an Abelian gauge field
takes  the form
\begin{equation}
\Gamma^{PV} = {ie\over 2\pi}\int d^{2}x\,\arctan \left(\tanh {\beta
M\over 2} \tan ({ea\over 2})\right) B(\vec{x})\label{fosco}
\end{equation}
However, because the choice of the background is very special, it
would seem that this may not represent the complete effective action
in a general background. In fact, in higher dimensions, such as $2+1$,
one also has to tackle with the question of the non-analyticity of the
thermal amplitudes which leads to a nonuniqueness of the effective
action  \cite{l1,l10}.

With these issues in mind, we have studied the parity violating part
of the four point function in $2+1$ dimensions at finite
temperature. The calculations are clearly extremely difficult and we
have evaluated the amplitudes at finite temperature by using the
method of forward scattering amplitudes  \cite{l11}.
Without going into details, let me summarize the results
here  \cite{l12}. First, the parity violating part of the box diagram is
nontrivial at zero temperature and comes from an effective action of
the form
\begin{equation}
\Gamma^{4}_{T=0} = - {e^{4}\over 64\pi M^{6}} \int
d^{3}x\,\epsilon^{\mu\nu\lambda}\,F_{\mu\nu}
(\partial^{\tau}F_{\tau\lambda}) F^{\rho\sigma}F_{\rho\sigma}
\end{equation}
This is Lorentz invariant and is invariant under both small and large
gauge transformations and is compatible with the Coleman-Hill theorem
 \cite{l13}.

At finite temperature, however, the amplitude is not manifestly
Lorentz invariant (because of the heat bath) and is non-analytic. We
have investigated the amplitude in two interesting limits. Namely, in
the long wave limit (all spatial momenta vanishing), the leading term
of the  amplitude, at high
temperature,  can be seen to come from an effective action of the form
\begin{equation}
\Gamma^{4}_{LW} = {e^{4}\over 512 MT}\int
d^{3}x\,\epsilon_{0ij}\,E_{i}(\partial_{t}^{-1}E_{j})
(\partial_{t}^{-1}E_{k}) (\partial_{t}^{-1}E_{k})
\end{equation}
where $\vec{E}$ represents the electric field. There are several things to
note here. First, this is an extensive action, be it
non-local. Second, it is manifestly large gauge invariant and finally,
the leading behavior at high temperature goes as ${1\over T}$.

In contrast, we can evaluate the amplitude in the static limit (all
energies vanishing) where we find the presence of both extensive as
well as non-extensive terms at high temperature. However, the
extensive terms are suppressed by powers of $T$ and the leading term
seems to come from an effective action of the form
\begin{equation}
\Gamma^{4}_{S} = {e^{4}\over 4\pi T^{2}}\left(\tanh {\beta M\over 2} -
\tanh^{3} {\beta M\over 2}\right) \int d^{3}x\,a^{3} B
\end{equation}
This coincides with the amplitude that will come from
eq. (\ref{fosco}) and has the leading behavior of ${1\over T^{3}}$ at
high temperature. Such a term is not invariant under a large gauge
transformation. However, we can now derive a large gauge Ward identity
for the leading part of the static action and the solution of the Ward
identity coincides with the form given in eq. (\ref{fosco}). This,
therefore, clarifies the meaning of the effective action in the
special background, namely, it represents the leading term in the
effective action in the static limit.

\section{Higher order corrections:}

Since we have calculated the box diagram at finite temperature, we can
also ask about possible higher loop corrections to the CS
coefficient. Let us note that, at zero temperature, there is a result
due to Coleman and Hill \cite{l13}, which says that in an Abelian theory, there
cannot be any correction to the CS coefficient beyond one loop. Their
result basically uses two simple assumptions, i) small gauge
invariance and ii) analyticity of the amplitudes in the momentum
space. Small gauge invariance is, of course, known to be true at
finite temperature. However, as we have pointed out earlier,
amplitudes become non-analytic in the momentum space at finite
temperature. Therefore, the second assumption of Coleman-Hill breaks
down at finite temperature and one may expect higher loop correction
to the CS coefficient at finite temperature.

We have explicitly computed the two loop correction to the CS
coefficient at finite temperature \cite{BDFR}. Parameterizing the
self-energy in a covariant gauge as
\begin{equation}
\Pi^{\mu\nu}(p,u) = \Pi^{\mu\nu}_{1}(p,u) + i \epsilon^{\mu\nu\lambda}
p_{\lambda} \Pi_{2}(p,u)
\end{equation}
where $u^{\mu}$ represents the velocity of the heat bath, we find
that, in the static limit, the two loop correction to the CS
coefficient at high temperatures takes the form
\begin{equation}
\Pi_{2}^{(2)}(0) = (2m - 3M)\,{e^{4}\over 192\pi^{2}T^{2}}\,\ln
{T\over m}
\end{equation}

This result explicitly shows that the Coleman-Hill result breaks down
at finite temperatures. Furthermore, the form of the correction is
interesting in that it diverges as $m\rightarrow 0$. This is the usual
manifestation of infrared divergence and shows that, although the zero
temperature theory is well defined in the limit of vanishing $m$,
there are infrared divergences at finite temperature. In addition,
since there is a two loop correction to the CS coefficient, one may
ask the structure of the effective action when higher order
corrections are taken into account. This can be determined from the
large gauge Ward identity that we have derived and it determines the
form of the parity violating effective action, satisfying large gauge
invariance, at any order to be
\begin{equation}
\Gamma^{PV} = \tan^{-1}\left(2\Gamma^{\prime}(0) \tan {ea\over
2}\right)\int d^{2}x\,B
\end{equation}
where $\Gamma^{\prime}(0)$ represents the correction to the CS
coefficient to that order.

\section{Non-Abelian theories:}

Our main interest was in the study of the question of large gauge
invariance at finite temperature in a non-Abelian theory. As we have
shown, the question of large gauge invariance is well understood in an
Abelian theory. However, not much is known about the non-Abelian
theory yet. To that extent, let us note that even the Coleman-Hill
result was derived for an Abelian theory. In this section, let me
describe briefly how the Coleman-Hill result can be generalized to
non-Abelian theories (at zero temperature).

There are several qualitative differences between the Abelian and the
non-Abelian theories. First, while the Abelian theory is well defined
even when the tree level CS coefficient vanishes, the infrared
divergences in the non-Abelian theory are severe if there is no tree
level CS term. Second, while the Abelian theory can be defined in any
gauge, the non-Abelian theory is well defined only in a select class
of infrared safe gauges such as the Landau gauge, the axial gauge
etc. Finally, in the Abelian theory, the CS coefficient is a gauge
independent quantity while in the non-Abelian theory, the CS
coefficient is gauge dependent. As a result, it is not clear how to
generalize the Coleman-Hill result in the non-Abelian case.

On the other hand, it is known from various arguments that, in a
non-Abelian theory, it is the ratio ${4\pi m\over g^{2}}$ which has a
physical meaning and, therefore, must be gauge independent. In fact, a
one loop calculation verifies that in all infrared safe gauges the one
loop correction to this ratio is given by
\begin{equation}
{4\pi m\over g^{2}} \rightarrow {4\pi m\over g^{2}} + N
\end{equation}
It is, therefore, meaningful to generalize the Coleman-Hill result for
this ratio. (Recall that it is also this ratio that needs to be
quantized for large gauge invariance to hold.)

This can indeed be done as follows \cite{BDF}. First, let us note that, although
the CS coefficient is gauge dependent in general, it takes on a
physical meaning in the axial gauge. This is because in the axial
gauge,
\begin{equation}
{4\pi m\over g^{2}} \rightarrow {4\pi m\over g^{2}}\,(1+\Pi_{2}(0))
\end{equation}
Since this ratio is physical, in this gauge $\Pi_{2}(0)$ does carry a
physical meaning. Using the Ward identities in the axial gauge, it is
straight forward to show that the CS coefficient does not receive any
corrections beyond one loop provided i) small aguge invariance holds
and ii) amplitudes are analytic in the momentum space. A consequence
of this is that the ratio ${4\pi m\over g^{2}}$ does not have any
correction beyond one loop in this gauge. However, this is a gauge
independent quantity and hence it holds in any gauge that this ratio
does not receive any correction beyond one loop. Thus, we understand
some features of the zero temperature non-Abelian theory which are
parallel to the Abelian theory and what remains is to understand
systematically if the issue of large gauge invariance also gets
resolved in a parallel manner. 

\section{Conclusion:}

In this talk, we have discussed the question of large gauge
invariance at finite temperature. We have discussed the resolution of
the problem in a simple $0+1$ dimensional model. We have derived the
Ward identity for large gauge invariance in this model. We have
analyzed the box diagram in $2+1$ dimensions and have obtained the
form of the effective action at zero temperature. We have also
obtained the amplitude as well as the quartic effective actions in the
long wave as well as static limits, at finite temperature. The LW
limit has only extensive terms in the action which goes as ${1\over
T}$ at high temperature and is invariant under large gauge
transformations. The leading term in the static action,
however, is non-extensive, goes as ${1\over T^{3}}$ at high
temperature and  coincides  with the effective action proposed earlier
for a restrictive gauge background. This action is not invariant under
large gauge transformations. However, using a large gauge Ward
identity, we can determine the full leading order action in the static
limit which coincides with the effective action obtained in a
restrictive gauge background. We have shown explicitly that higher
loop corrections to the CS coefficient do not vanish at finite
temperature. This violation of the Coleman-Hill result is a
consequence of the fact that one of their assumptions (namely,
analyticity of the amplitudes) breaks down at finite temperature. We
have also extended the result of Coleman-Hill to the case of
non-Abelian gauge theories.

It is a plaesure to thank the organizers for their
hospitality. I have learnt a lot from all of my collaborators in the
field, namely, K. Babu, J. Barcelos-Neto, F. Brandt, A. J. da Silva,
G. Dunne, J. Frenkel, P. Panigrahi, K. Rao and J. C. Taylor, and I am
grateful to all of them. This work was supported in
part by the U.S. Dept. of Energy Grant  DE-FG 02-91ER40685.

\end{document}